\documentclass[secnumarabic,amssymb, nobibnotes, aps, prd,showkeys]{revtex4-2}

\usepackage{hyperref}
\usepackage{graphicx}
\usepackage{caption}
\usepackage{amsmath}
\usepackage{amssymb}
\usepackage{graphicx}

\begin{document}

\title{Hidden-charm and hidden-beauty hybrid mesons}%

	\author{Raghavendra Kaushal}%
\email[Raghavendra Kaushal: ]{raghavendra1.mitmpl2022@learner.manipal.edu}
\affiliation{Manipal Institute of Technology, Manipal Academy of Higher Education, Manipal, India.}
\author{Bhaghyesh}%
\email[Bhaghyesh: ]{bhaghyesh.mit@manipal.edu}
\affiliation{Manipal Institute of Technology, Manipal Academy of Higher Education, Manipal, India.}
\begin{abstract}
	The mass spectra of $\Pi_u$, $\Sigma_u^-$, $\Pi_g$, $\Delta_g$, $\Sigma_g^-$, $\Delta_u$, and $\Sigma_u^+$ hybrid charmonium and hybrid bottomonium states are computed using a non-relativistic potential model. The hybrid potential is modelled as a linear combination of quarkonium potential and gluonic excited potential. The gluonic excited potential is computed by fitting the potential difference between hybrid potential and quarkonium potential obtained from the lattice quantum chromodynamics. Experimental masses and $J^{PC}$ values of $\Psi(4230)$, $\Psi(4500)$, $X(4500)$, and $\Psi(4660)$ are compatible with hybrid charmonium states, and $\Upsilon(10753)$ and $\Upsilon(11050)$ are compatible with hybrid bottomonium states obtained from our model, respectively. 
\end{abstract}


\maketitle

\section{Introduction}
Quantum Chromodynamics (QCD) defines hadrons as color singlet particles, that are bound states of quarks (and/or antiquarks). The color interaction binds these quarks (and/or antiquarks) mediated through SU(3) gauge boson called gluons. There can be a number of combinations in which the quarks ($q$) or antiquarks ($\bar{q}$) can combine to give color singlet states, the simplest being $q\bar{q}$ described as mesons and $qqq$ described as baryons \cite{1} (collectively called as conventional hadrons). Almost all the hadrons discovered until early 2000s were classified either as mesons or baryons. However, according to the theory of QCD, there can be more diverse combinations that can lead to color singlet states like $qq\bar{q}\bar{q}$, $qqqq\bar{q}$, $qqq\bar{q}\bar{q}\bar{q}$, and so on, as proposed by Gell-Mann \cite{1}. The hadrons that cannot be classified either as mesons or baryons are referred as an exotic hadrons. There are two ways in which one can identify exotic hadrons. Exotic hadrons can either have quantum numbers that cannot be reproduced by considering conventional hadronic states or can exhibit properties during its production or decays that is inconsistent with the conventional hadrons \cite{3}. Various theoretical interpretations exists for exotic states. Exotic states can be a compact multiquark states, where quarks are bound through color interaction in a correlated way \cite{2,8,12}. It can be a molecular state, where the system is described as a compound system of color neutral particle, like a bound state of meson-meson or baryon-meson \cite{2,8,13}. It can be a hadroquarkonium state, where the system is described by a light-quark cloud surrounding the color-singlet quarkonium core \cite{2,8,14}. It can be a hybrid hadron state, where the gluonic field which binds the quarks (or antiquarks) might be excited to form a state \cite{3}. It can be glueballs, where gluons, unlike photons, can self-interact to form a state, entirely made up of gluons \cite{2}. In 2003, Belle Collaboration \cite{4} discovered a particle, named as $X(3872)$ with mass $3872.0\pm0.6\pm0.5$ MeV and decay width, $\Gamma < 2.3$ MeV with the confidence level of $90\%$. They observed that in the $\gamma \chi_{c1}$ decay channel, there was no strong signal, as was supposed to be present for a conventional charmonium state in this mass range and hence suggested this state to be an exotic state. This marked the first experimental discovery of an exotic state. There are various interpretations for $X(3872)$ as a molecular state \cite{6}, a tetraquark state \cite{5}, superposition of conventional charmonium and molecular state \cite{7}, and so on. In 2005, BaBar Collaboration \cite{17} discovered a particle while analyzing the $e^+ e^- \rightarrow \gamma_{ISR} \pi^+ \pi^- J/\psi$, an initial-state radiation event and named as $Y(4260)$, with mass around $4260$ MeV. The interesting feature of this particle that differentiates this from conventional charmonium state is that it decays into hidden charm meson, $Y(4260) \rightarrow J/ \psi \pi^+ \pi^-$ \cite{17} instead of open charm meson pairs, even after being lying above the open-charm threshold. There are various interpretations for this particle as hybrid charmonium meson \cite{19}, hadronic molecule and compact tetraquark states \cite{20}. The first charged exotic state was discovered in 2007 by Belle Collaboration \cite{18} with mass $4433 \pm 4$ (stat) $\pm 2$ (syst) MeV and decay width of $45^{+18}_{-13}$ (stat) $\, ^{+30}_{-13}$ (syst) MeV, while analyzing the decay $B \rightarrow K \pi^+ \psi'$. This particle is named as $Z^+(4430)$ and is believed to be a strong candidate for a first radially excited compact tetraquark state \cite{21,22}. Since $Z^+(4430)$ is believed to be an excited state, there must exist a ground state tetraquark, that must decay into $\pi^+ \pi^- J/\psi$. This particle was observed simultaneously by Belle \cite{24} and BES III collaboration \cite{23} in 2013, while analyzing the decay process, $e^+ e^- \rightarrow \pi^+ \pi^- J/ \psi$. This particle was named as $Z^+(3900)$, with observed mass of $3894.5\pm6.6\pm4.5$ MeV and decay width of $63\pm25\pm26$ MeV by Belle collaboration \cite{24} and observed mass of $3899.0\pm3.6\pm4.9$ MeV and decay width of $46\pm10\pm20$ MeV, respectively by BES III collaboration \cite{23}.  In 2011, Belle collaboration \cite{25} reported two resonance structures $Z_b(10610)$ with mass $10607.2\pm2.0$ MeV and decay width of $18.4\pm2.4$ MeV and $Z_b(10650)$ with mass $10652.2\pm1.5$ MeV and decay width of $11.5\pm2.2$ MeV, respectively. These were the first exotic states discovered that consisted of bottom quarks. The first widely accepted pentaquark states $P_c(4380)^+$ and $P_c(4450)^+$ were observed in the decay channel, $\Lambda_{b}^{0} \rightarrow J/ \psi K^- p$ by the LHCb collaboration \cite{26} in 2015. Presently, we have more than 50 exotic hadron candidates observed experimentally, which has been illustrated innovatively in Fig. 2 of Ref.\cite{8}. LHCb collaboration played a major role in discovering most of the observed exotic states \cite{8,9}. However, the experimental study of exotic hadrons are still in the developing stage. The $e^+e^-$, $p\bar{p}$, $pp$, and relativistic heavy ion collision setup available at LHCb collaboration provides a unique oppurtunity for the discovery and analysis of exotic states \cite{27}. Belle II collaboration's Super KEKB collider has a unique feature of operating at the centre of mass energy close to $\Upsilon(4S)$ ($\approx 10$ GeV), offering high probability for the discovery of exotic hadrons \cite{28}. Hence it becomes important to explore exotic hadrons theoretically.

In this work, we explore hybrid mesons among the other classes of exotic hadrons. Hybrid mesons can either be treated as a three body system consisting of a quark, an antiquark, and a constituent gluon or as a two body system consisting of a quark and an antiquark with excited gluonic field \cite{9a}. In the fundamental color representation, gluons will be described by color octet ($\mathbf{8}$), quarks by color triplet ($\mathbf{3}$), and antiquark by color antitriplet ($\mathbf{\bar{3}}$) state respectively. The combination of quark and antiquark gives $\mathbf{3} \otimes \mathbf{\bar{3}} = \mathbf{1} \oplus \mathbf{8}$. In the first case, where hybrid meson is considered as a three body system, the combination of quark and antiquark will be considered in color octet state ($\mathbf{8}$), and the combination with a constituent gluon, $\mathbf{8} \otimes \mathbf{8} = \mathbf{27} \oplus \mathbf{10} \oplus \mathbf{10} \oplus \mathbf{8} \oplus \mathbf{8} \oplus \mathbf{1}$, results in a hybrid meson in color singlet state ($\mathbf{1}$). In the second case, where hybrid meson is considered as a two body problem with gluonic excitation, the combination of quark and antiquark will be considered in a color singlet ($\mathbf{1}$) state. It is observed that the two treatments of hybrid mesons are equivalent to each other \cite{10,11}. In this work, we consider hybrid meson to be a bound state of quark and antiquark with an excited gluonic field. One of the successful framework employed to study the gluonic excitation potential is the Born-Oppenheimer (B-O) approximation \cite{15} for $Q\bar{Q}$ meson system. In this framework, hybrid meson is considered as a diatomic molecule, where $Q$ and $\bar{Q}$ will correspond to the nucleus of the two atoms and the gluonic field will correspond to the field of the electron in the molecule. First the quark and antiquark will be considered as a static color sources and the excited gluonic field energies are determined as a function of interquark separation $r$. The quark motion will then be considered by solving the Schrodinger equation \cite{15}. In this work, we will be determining the mass spectra of various hidden-charm hybrid mesons ($c\bar{c}$ with gluonic excitation, also called hybrid charmonium) and hidden-beauty hybrid mesons ($b\bar{b}$ with gluonic excitation, also called hybrid bottomonium) using phenomenological potential model under the Born-Oppenheimer approximation. The paper is structured as follows. In section \ref{S2}, we have given the theoretical framework where we discuss the conserved quantum numbers that characterize hybrid mesons, construction of angular momentum, and the potential model used to compute the mass spectra. In section \ref{S3}, we discuss our results, and in section \ref{S4}, we summarize our work.

\section{Theoretical Framework: \label{S2}}
\subsection{Quantum Numbers}
Hybrid mesons are characterized by three quantum numbers ($\Lambda$, $\eta$, and $\epsilon$) that are conserved. In the B-O formalism, a hybrid state is denoted as $\Lambda_{\eta}^{\epsilon}$. These quantum numbers are defined as follows \cite{3,15,16}:

i) $\lambda$ corresponds to the eigenvalue of $\mathbf{\hat{r}} \cdot \mathbf{J_g}$. Here $\mathbf{J_g}$ is the total angular momentum of the gluonic field and $\mathbf{r}=r \mathbf{\hat{r}}$ is the interquark seperation vector. $\lambda$ can have values $0, \pm1, \pm2,\ldots$. The absolute value of  $\lambda$, represented by $\Lambda$ (i.e, $\Lambda = |\lambda|$) will be used to define the hybrid state. $\Lambda=0$ is represented as $\Sigma$, $\Lambda=1$ as $\Pi$, $\Lambda=2$ as $\Delta$, and so on respectively.

ii) $\eta$ corresponds to the eigenvalue of the product $C_gP_g$, where $C_g$ and $P_g$ are the charge conjugation operator and parity operator for the gluonic field respectively. $\eta$ can take values $-1$ and $+1$. $\eta=-1$ is represented as $u$ and $\eta=+1$ as $g$, respectively.

iii) For $\lambda=0$, $\epsilon$ corresponds to the eigen value of $R_g$, the operator that reflects the gluonic field across the plane that contains $\bar{Q}$ and $Q$ sources. $\epsilon$ can have possible values of $+1$ and $-1$, represented as $+$ and $-$, respectively. For $\lambda \neq 0$, the hybrid state will be degenerate for $\epsilon=-$ and $+$.

\subsection{Angular Momentum}
The angular momentum $\mathbf{J}$ of a hybrid meson is constructed as follows \cite{3,15,16}. First the total angular momentum of gluonic field $\mathbf{J_g}$ couples with the relative orbital angular momentum $\mathbf{L_{Q\bar{Q}}}$ between $Q$ and $\bar{Q}$ to obtain an angular momentum $\mathbf{L}$,
\begin{eqnarray} \label{E1}
	\mathbf{L_{Q\bar{Q}}}+\mathbf{J_g}=\mathbf{L}
\end{eqnarray}
The angular momentum $\mathbf{L}$ then couples with resultant spin $\mathbf{S}$ of $Q$ and $\bar{Q}$ to give total angular momentum $\mathbf{J}$,
\begin{eqnarray}
	\mathbf{L}+\mathbf{S}=\mathbf{J}
\end{eqnarray}
L must satisfy the condition $L \geq \Lambda$ \cite{3,15,16}. This can be shown as follows.
\begin{eqnarray}
	\mathbf{\hat{r}} \cdot \mathbf{L} = \mathbf{\hat{r}} \cdot \left(\mathbf{L_{Q\bar{Q}}}+\mathbf{J_g}\right)
\end{eqnarray}
Since $\mathbf{L_{Q\bar{Q}}}=\mathbf{r} \times \mathbf{p}$, $\mathbf{L_{Q\bar{Q}}}$ will be perpendicular to $\mathbf{\hat{r}}$, resulting in $\mathbf{\hat{r} \cdot L_{Q\bar{Q}}}=0$. Thus we have,
\begin{eqnarray} \label{E4}
	\mathbf{\hat{r} \cdot L}=\mathbf{\hat{r} \cdot J_g}=\lambda
\end{eqnarray}
The projection of angular momentum $\mathbf{L}$ on any axis corresponds magnetic quantum number $m_L$ and its magnitude cannot exceed $L$ i.e., $|m_L| \leq L$. This $m_L$ corresponds to $\lambda$ in Equation (\ref{E4}) and hence we have the condition $L\geq\Lambda$. The parity (P) and charge conjugation (C) of hybrid mesons are described in terms $L$, $S$, $\Lambda$, $\epsilon$, and $\eta$ \cite{3,15,16},  
\begin{eqnarray}
	P=\epsilon (-1)^{L+\Lambda +1}, \quad
	C=\epsilon \eta (-1)^{L+\Lambda +S}
\end{eqnarray}
In addition to the $J^{PC}$ values observed in conventional quarkonium states, due to the excited gluonic fields, hybrid mesons can have exotic $J^{PC}$ values, not observed in conventional quarkonium states. The possible exotic $J^{PC}$ values observed in hybrid mesons are $\mathbf{0^{--}}$, $\mathbf{0^{+-}}$, $\mathbf{1^{-+}}$, $\mathbf{2^{+-}}$, $\mathbf{3^{-+}}$, and so on. In the upcoming sections, we represent these exotic $J^{PC}$ values of hybrid mesons using bold fonts to distinguish from the conventional $J^{PC}$ values.
\subsection{Mass Spectra}

The mass spectra of hybrid mesons are computed by solving the time-independent Schrodinger equation given by \cite{3,15,16},
\begin{eqnarray}
		\left[\frac{1}{2\mu}\left(-\frac{d^2}{d r^2}+\frac{\langle L_{Q\bar{Q}}^2\rangle }{r^2}\right)+V_{\Gamma}(r)\right]u(r)=Eu(r).
\end{eqnarray}
The expectation value of $L_{Q\bar{Q}}^2$ is obtained as follows. From Equation (\ref{E1}), we have,
\begin{eqnarray}
	L_{Q\bar{Q}}^2=L^2-2\mathbf{L\cdot J_g}+J_g^2
\end{eqnarray}
Lets define the projections,
\begin{eqnarray}
	\mathbf{\hat{r} \cdot L}=L_r, \quad
	\mathbf{\hat{r}\cdot J_g}=J_{gr}
\end{eqnarray}
Now the dot product $\mathbf{L\cdot J_g}$ becomes \cite{3},
\begin{eqnarray} \label{E8}
	\mathbf{L\cdot J_g}=L_rJ_{gr}+\frac{1}{2}\left(L_+J_{g-}+L_-J_{g+}\right)
\end{eqnarray}
Using Equation (\ref{E4}) we can write $L_rJ_{gr}=\lambda^2=\Lambda^2$, and hence the Schrodinger equation becomes,
\begin{eqnarray} \label{E9}
	\left[\frac{1}{2\mu}\left(-\frac{d^2}{d r^2}+\frac{\langle L^2-2\Lambda^2+J_g^2-\left( L_+J_{g-}+L_-J_{g+} \right) \rangle }{r^2}\right)+V_{\Gamma}(r)\right]u(r)=Eu(r).
	\end{eqnarray}
In Equation (\ref{E9}), the final term in the expression for $\langle L_{Q\bar{Q}}^2 \rangle$ is a linear combination of the operators $J_{g-}$ and $J_{g+}$. These operators couple the wave function to states whose $\Lambda$ differs by one unit within a multichannel Schrödinger framework. If one assumes a single-channel case, such coupling terms can be neglected \cite{3}. Consequently, when the wave function is an eigenstate of $L^2$ with angular momentum
$L$, the expectation value of $L_{Q\bar{Q}}^2$ reduces to,
\begin{eqnarray}
	\langle L_{Q\bar{Q}}^2 \rangle = L(L+1)-2 \Lambda^2+ J_g(J_g+1)
\end{eqnarray}
As a result, the Schrodinger equation for hybrid mesons becomes,
\begin{eqnarray}
	\left[\frac{1}{2\mu}\left(-\frac{d^2}{d r^2}+\frac{L(L+1)-2 \Lambda^2 + J_g(J_g+1) }{r^2}\right)+V_{\Gamma}(r)\right]u(r)=Eu(r).
\end{eqnarray}
The gluon angular momentum, $J_g$ is 1 for $\Pi_u$, $\Pi_g$, and $\Sigma_u^-$, and 2 for $\Sigma_u^+$, $\Sigma_g^-$, $\Delta_u$, and $\Delta_g$ \cite{16}. $V_{\Gamma}(r)$ is the hybrid potential for specific hybrid state $\Lambda_{\eta}^{\epsilon}$. In our work, we consider $V_{\Gamma}(r)$ as a linear combination of quark-antiquark potential ($V_{Q\bar{Q}}(r)$) and the gluonic excited potential ($V_g(r)$), i.e.,
\begin{eqnarray}
	V_{\Gamma}(r)=V_{Q\bar{Q}}(r)+V_g(r)
\end{eqnarray}
The quark-antiquark potential that we consider in this work is given by,
\begin{eqnarray} \label{E13}
		V_{Q\bar{Q}}(r)=-\frac{4}{3} \frac{\alpha_{s}}{r}+\frac{b(1-e^{-\nu r})}{\nu}+\left(\frac{1}{m_1} +\frac{1}{m_2}\right)\left(\frac{-9\alpha^{2}}{8r^2}+C\ln(ar)\right)\nonumber\\
	+\frac{32\pi\alpha_s}{9m_1m_2}\left(\frac{\sigma}{\sqrt{\pi}}\right)^3e^{-\sigma^2 r^2} \mathbf{S_1}\cdot\mathbf{S_2}.
\end{eqnarray}
We have utilized this potential in our previous works where we have determined the mass spectra of charmed (charmonium and triply charmed baryons) \cite{K1} and bottom (bottomonium and triply bottom baryons) hadrons \cite{K2}. The potential model parameters of Equation (\ref{E13}) for charmed and bottom hadrons are given in Table (\ref{T0}). In this work, we will be computing the mass spectra of $\Pi_u$, $\Sigma_u^-$, $\Pi_g$, $\Delta_g$, $\Sigma_g^-$, $\Delta_u$, and $\Sigma_u^+$ hybrid states. The gluonic excitation potential $V_g(r)$ for each of these hybrid configurations are obtained by fitting the difference between the corresponding hybrid potential and the ground state (quarkonium) potential obtained from the lattice studies of Ref. \cite{16}. We find that this difference between the potential of all the hybrid states and the ground state (quarkonium) from the lattice study \cite{16} to be of the form $V_g (r)=X+Ye^{-Zr}$, where $X$, $Y$, and $Z$ are the phenomenological constants. $X$ indicates the asymptotic value of the gluonic excitation potential when the interquark separation is large (i.e., as $r \rightarrow \infty$), whereas $Y$ determines the strength of the radially dependent contribution of the excited gluonic field, and $Z$ determines the rate at which the radial contribution decreases with the increase in the interquark separation. The numerical values of these parameters are given in Table (\ref{T1}) for different hybrid states. As $V_g (r)$ is the gluonic excitation potential, it does not depend on the quark flavors and hence, the fitted parameters remain the same for both hybrid charmonium and hybrid bottomonium mesons. In this work, we have not included spin-orbit and tensor interactions because we found that the spin-spin interaction, dominant across the spin dependent interactions did not produce any difference in mass for $S=0$ and $S=1$ hybrid meson states as can be observed in Tables (\ref{C1}-\ref{B7}). Also, in principle, the hybrid quarkonium state can mix with conventional quarkonium states with identical quantum numbers, as can be observed in Ref. \cite{37} . However, this analysis is beyond the scope of the present work. Also the authors in Ref. \cite{37} have used the Born-Oppenheimer effective field theory (BOEFT) in which they take into account $\Pi_u$ and $\Sigma_u^-$ non-adiabatic mixing through coupled Schrodinger wave equation. In that formalism, the potentials corresponding to $\Pi_u$ and $\Sigma_u^-$ originate from the same gluelump configuration resulting in a nearly degenerate states at short distances. However, in our analysis as well as in Ref. \cite{3,16}, $\Pi_u$ and $\Sigma_u^-$ are treated independently, without the non-adiabatic mixings. 

\begin{table}
	\caption{Potential model parameters of $V_{Q\bar{Q}}(r)$.}
	\begin{ruledtabular}
		\begin{tabular}{@{}ccccccc} \label{T0}
			&$m_1=m_2$ (GeV)&$b$ (GeV$^2$)&$\nu$ (GeV)&$C$ (GeV)&$a$ (GeV)&$\sigma$ (GeV)\\
			\hline
			Charmed hadrons&$1.397$&$0.212$&$0.072$&$-0.148$&$0.215$&$3.520$\\
			Bottom hadrons&$4.680$&$0.241$&$0.078$&$0.100$&$0.430$&$3.920$\\
		\end{tabular}
	\end{ruledtabular}
\end{table}

\begin{table}
	\caption{Model Parameters of $V_g(r)$ for different hybrid states}
	\begin{ruledtabular}
		\begin{tabular}{@{}cccc}\label{T1}
			$\Lambda_{\eta}^{\epsilon}$&$X (GeV)$&$Y (GeV)$&$Z (GeV)$\\
			\hline
			$\Pi_u$&$0.384$&$1.440$&$0.455$\\
			$\Sigma_u^-$&$0.903$&$0.944$&$0.700$\\
			$\Pi_g$&$0.227$&$1.800$&$0.181$\\
			$\Delta_g$&$0.647$&$1.657$&$0.340$\\
			$\Sigma_g^-$&$1.076$&$1.306$&$0.505$\\
			$\Delta_u$&$0.823$&$1.798$&$0.260$\\
			$\Sigma_u^+$&$0.973$&$1.656$&$0.361$\\
		\end{tabular}
	\end{ruledtabular}
\end{table}

\section{Results and Discussions: \label{S3}}
\begin{figure}
	\centering
	\includegraphics[width=1\linewidth]{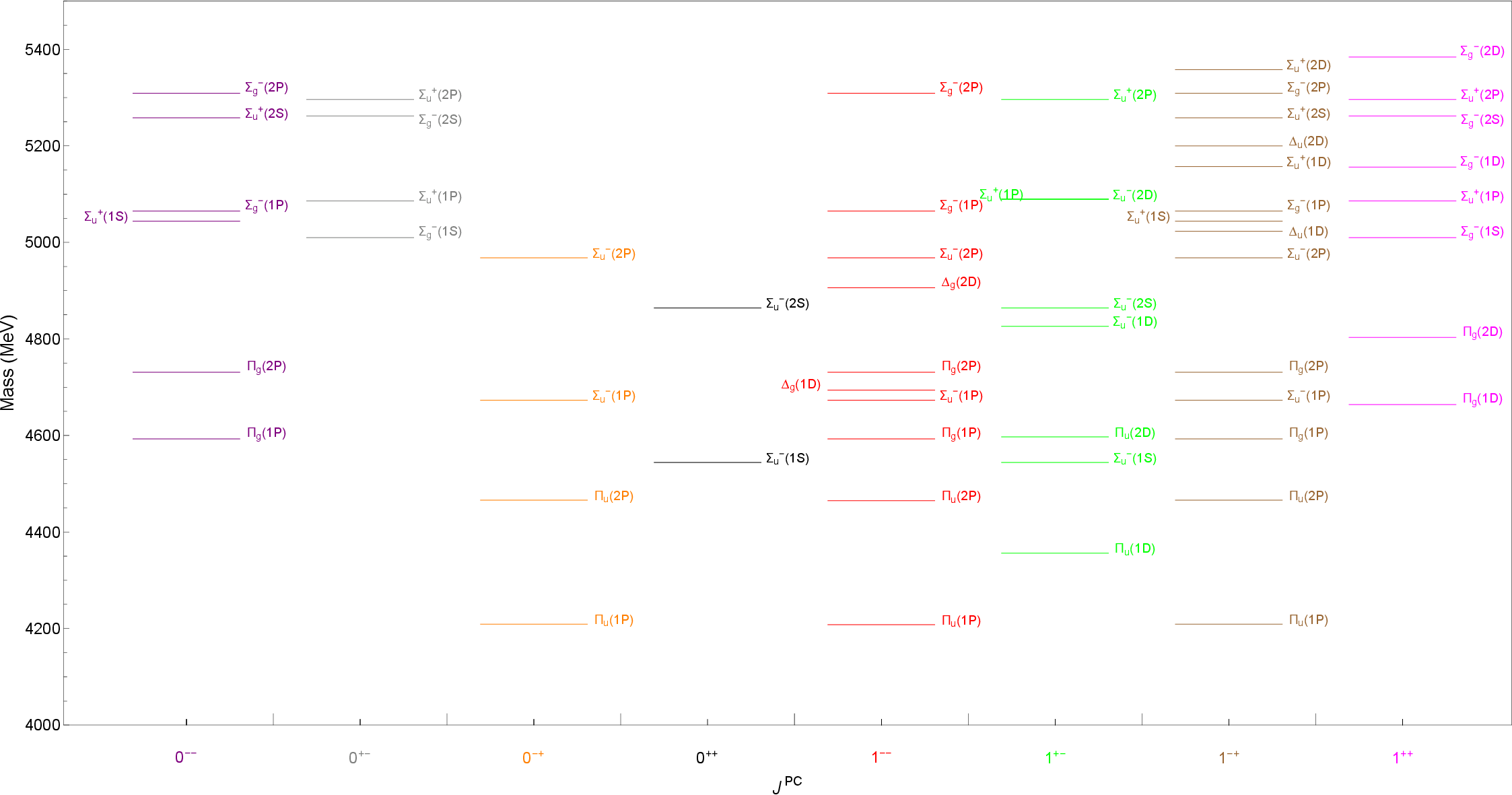} 
	\caption{Mass spectra of hybrid charmonium states with $J=0,1$}
	\label{F1}
\end{figure}

\begin{figure}
	\centering
	\includegraphics[width=1\linewidth]{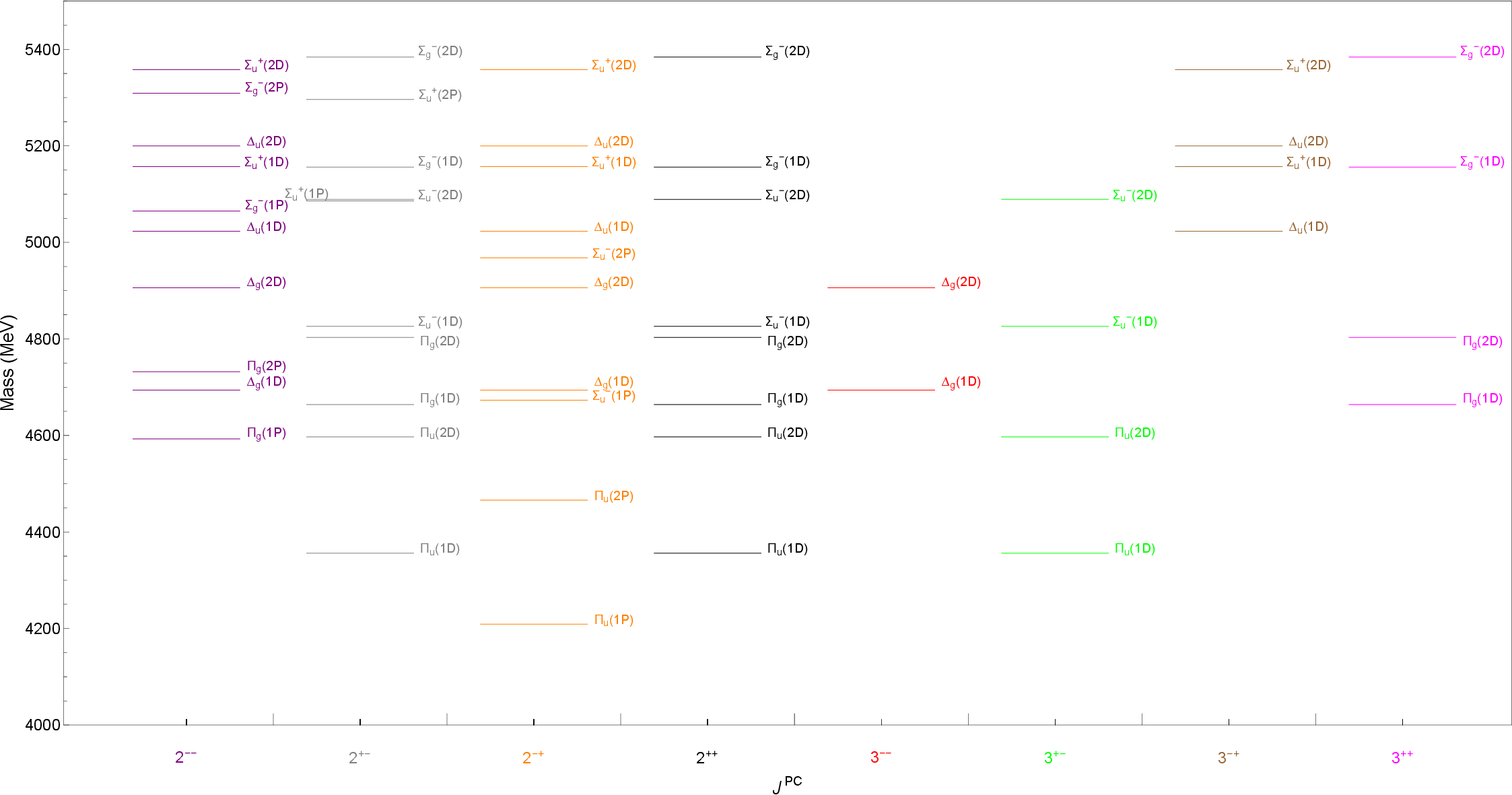} 
	\caption{Mass spectra of hybrid charmonium states with $J=2,3$}
	\label{F2}
\end{figure}

We have presented the mass spectra of various hidden-charm hybrid mesons in Tables (\ref{C1}-\ref{C7}) and is schematically presented in Figures (\ref{F1}-\ref{F2}). The ground state hybrid charmonium corresponds to the $\Pi_u$ hybrid state with predicted mass of $4208$ MeV from our model. The ground state charmonium mass predicted in our previous work \cite{K1} is $2981$ MeV, around $1227$ MeV less than the ground state mass of hybrid charmonium state. This observation is consistent with the Lattice study \cite{35} which predicts this difference to be around $1200$ - $1300$ MeV. The lowest lying charmonium hybrid with exotic quantum number, $J^{PC}=\mathbf{1^{-+}}$ has a predicted mass of $4.19$ GeV from the flux tube model \cite{38}, $4.02\pm0.08$ GeV from the QCD sum rules \cite{39}, $4.184(0.006)$ GeV \cite{16} from the parameterized lattice results, and $4.011$ GeV from the BOEFT \cite{37}. We predict this mass to be $4.209$ GeV, consistent with other theoretical approaches. Recently, BES III collaboration \cite{31} observed three vector charmonium like resonant structures $R_1$, $R_2$, and $R_3$ with $J^{PC}=1^{--}$ (described as $Y$ states), with masses of $4223.6^{+3.6+2.6}_{-3.7-2.9}$ MeV$/c^2$, $4327.4^{20.1+10.7}_{-18.8-9.3}$ MeV$/c^2$, and $4467.4^{+7.2+3.2}_{-5.4-2.7}$ MeV$/c^2$ respectively and decay widths of $58.5^{+10.8+6.7}_{-11.4-6.5}$ MeV, $244.1^{+34.0+24.2}_{-27.1-18.3}$ MeV, and $62.8^{+19.2+9.9}_{-14.4-7.0}$ MeV respectively, by measuring the cross section of $e^+e^- \rightarrow \pi^+\pi^-h_c$. The $Y$ states decaying into $h_c$ is of particular interest because this decay is suppressed due to the heavy quark spin symmetry, suggesting that these $Y$ states are exotic in nature favoring the hybrid interpretation. They infer that their results for $R_1$ is consistent with $\psi(4230)$ and $R_2$ with $\psi(4360)$ listed in PDG \cite{PDG}, and $R_3$ with $\psi(4500)$ observed in their previous works \cite{32,33} respectively. Within our model, we assign $\psi(4230)$ with the PDG average mass of $4221.7\pm2.5$ MeV \cite{PDG} as a $\Pi_u$ hybrid charmonium state with $L=1$, $S=0$, $n=1$, and $J^{PC}=1^{--}$ with the predicted mass of $4208$ MeV. $\psi(4500)$ with the observed mass of $4467.4^{+7.2+3.2}_{-5.4-2.7}$ MeV$/c^2$ \cite{31} is assigned as a $\Pi_u$ hybrid charmonium state with $L=1$, $S=0$, $n=2$, and $J^{PC}=1^{--}$ with the predicted mass of $4465$ MeV. However, the experimental results on $\psi(4500)$ is limited and hence it is not listed in PDG \cite{PDG}. In our previous work \cite{K1}, we have assigned $\psi(4360)$ as $3\, {}^3D_1$ charmonium state. Eventhough hybrid interpretation for this state is possible, our model does not predict a hybrid charmonium state with $J^{PC}=1^{--}$, with the mass compatible with the observed mass of $\psi(4360)$. Hence, we continue our interpretation of $\psi(4360)$ state as the conventional $3\, {}^3D_1$ charmonium state. LHCb collaboration \cite{34} confirmed the existence of an exotic resonance $X(4500)$ with $J^{PC}=0^{++}$ and mass of $4512.5^{+6.0}_{-6.2}\pm3.0$ MeV by analyzing the diffractive $pp$ collisions through $J/\psi \phi$ final state. The observed mass and $J^{PC}$ values of $X(4500)$ is compatible with $\Sigma_u^-$ hybrid charmonium state with $L=0$, $S=0$, $n=1$, and $J^{PC}=0^{++}$, with the predicted mass of $4544$ MeV. The PDG average mass of $X(4500)$ is $4484\pm12$ MeV \cite{PDG}. $\psi(4660)$, previously known as $Y(4660)$, with mass $4623\pm10$ MeV and $J^{PC}=1^{--}$, listed as charmonium-like state in PDG \cite{PDG}, is compatible with  $\Sigma_u^-$ hybrid charmonium state with $L=1$, $S=0$, $n=1$, and $J^{PC}=1^{--}$, with the predicted mass of $4673$ MeV. The lower state mass of tensor hybrid charmonium state corresponding to $2^{-+}$ and $2^{++}$ predicted by QCD sum rules in Ref. \cite{29} is $4.16\pm0.14$ GeV and $4.5\pm0.1$ GeV respectively and in Ref. \cite{30}, it is $4.04\pm0.23$ GeV and $4.45\pm0.27$ GeV respectively. We predict these masses to be $4.209$ GeV and $4.356$ GeV respectively, which is compatible with those predicted by LQCD \cite{30} and QCD sum rules \cite{29}. We predict the masses of the lowest hybrid charmonium states with exotic quantum numbers $\mathbf{2^{+-}}$ as $4356$ MeV, $\mathbf{0^{--}}$ as $4593$ MeV, $\mathbf{0^{+-}}$ as $5010$ MeV, and $\mathbf{3^{-+}}$ as $5023$ MeV respectively.

\begin{figure}
	\centering
	\includegraphics[width=1\linewidth]{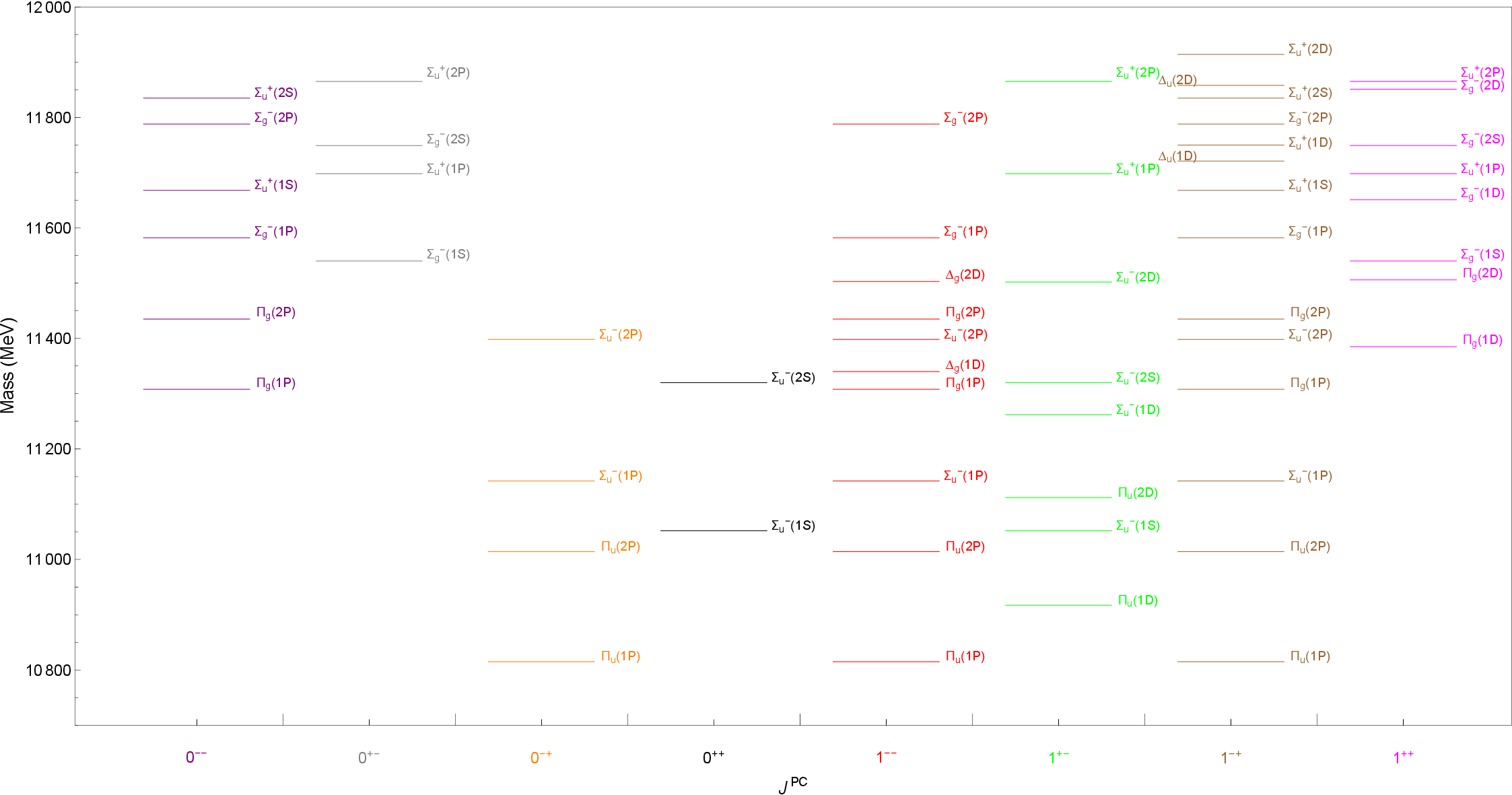} 
	\caption{Mass spectra of hybrid bottomonium states with $J=0,1$}
	\label{F3}
\end{figure}

\begin{figure}
	\centering
	\includegraphics[width=1\linewidth]{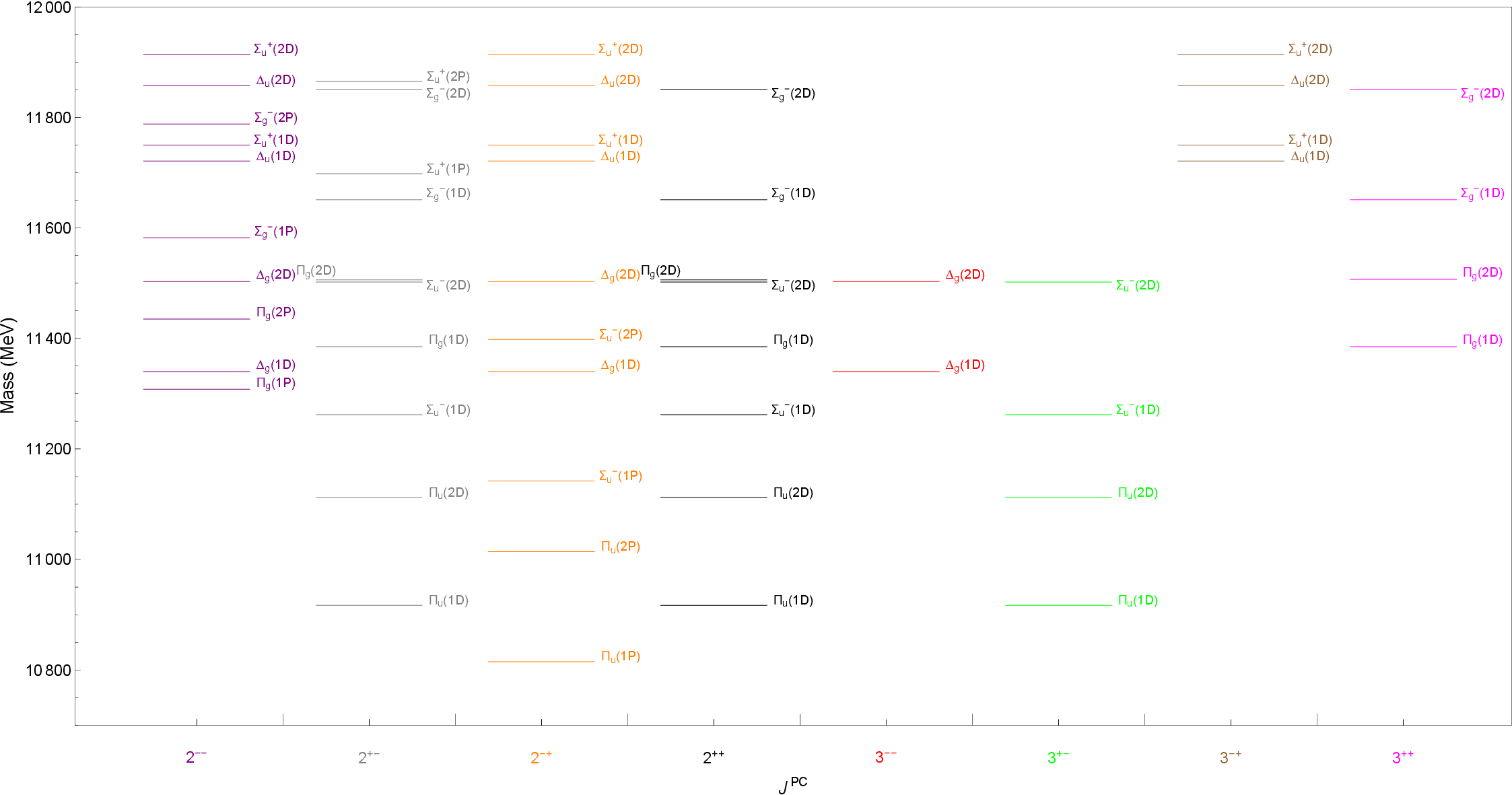} 
	\caption{Mass spectra of hybrid bottomonium states with $J=2,3$}
	\label{F4}
\end{figure}

We have presented the mass spectra of various hiden-beauty hybrid mesons in Tables (\ref{B1}-\ref{B7}) and in Figures (\ref{F3}-\ref{F4}). The ground state hybrid bottomonium state corresponds to the $\Pi_u$ hybrid state with predicted mass of $10815$ MeV from our model. The ground state bottomonium mass predicted in our previous work \cite{K2} is $9401$ MeV, around $1414$ MeV less than the ground state mass of hybrid bottomonium state. This observation is consistent with the Lattice study \cite{36} which predicts this difference to be around $1500$ MeV. The lowest lying exotic bottomonium hybrid with $J^{PC}=\mathbf{1^{-+}}$ has a predicted mass of $10.79$ GeV from the flux tube model \cite{38} and $10.83$ GeV from the QCD string approach \cite{40}. Our model predicts this mass to be $10.815$ GeV. In our previous work \cite{K2}, we predicted the mass of $\Upsilon(1S)$ to be $9458$ MeV and the mass difference between the lowest lying $\mathbf{1^{-+}}$ and $\Upsilon(1S)$ from our model is $1357$ MeV. Ref. \cite{40} predicts this difference to be $1430$ MeV. We predict the masses of the lowest hybrid bottomonium states with exotic quantum numbers $\mathbf{2^{+-}}$ as $10917$ MeV, $\mathbf{0^{--}}$ as $11308$ MeV, $\mathbf{0^{+-}}$ as $11540$ MeV, and $\mathbf{3^{-+}}$ as $11721$ MeV respectively. In our previous work \cite{K2}, we assigned $\Upsilon(10753)$ as $3\, {}^3D_1$ bottomonium state. A recent study by Belle II collaboration \cite{41} on $\Upsilon(10753) \rightarrow \gamma \chi_{bJ}$ radiative transitions failed to observe a significant signal, disfavoring the interpretation of $\Upsilon(10753)$ as a pure $D$- wave bottomonium state as this radiative transition is expected to have a relatively strong signal for a conventional $D$- wave bottomonium state. However, its too early to completely exclude the interpretation of $\Upsilon(10753)$ as a conventional bottomonium state. We can observe from our model that $\Pi_u$ hybrid bottomonium state with $L=1$, $S=0$, $n=1$, and $J^{PC}=1^{--}$ has a predicted mass of $10815$ MeV, close to the observed mass of $\Upsilon(10753)$ i.e., $10756.6\pm2.8$ MeV \cite{PDG}. The interpretation of $\Upsilon(10753)$ as a hybrid bottomonium state provides an interesting alternative to conventional bottomonium state. This dual interpretation for $\Upsilon(10753)$ is consistent with a recent BOEFT study \cite{42} where $\Upsilon(10753)$ is identified either as a $3\, {}^3D_1$ conventional bottomonium state or as a hybrid bottomonium state. More experimental results are required to validate the internal structure of $\Upsilon(10753)$. The authors in Ref. \cite{43} have assumed $\Upsilon(10753)$ and $\Upsilon(11020)$ as the ground state and the first radially excited hybrid bottomonium respectively with $J^{PC}=1^{--}$ and have calculated the transition decay widths of these particles into conventional bottomonium states using pNRQCD approach. By comparing their results with the experimental transition decay widths, they conclude that $\Upsilon(10753)$ and $\Upsilon(11020)$ are the two lowest lying hybrid bottomonium states. From our model, if we interpret $\Upsilon(10753)$ as a $\Pi_u$ hybrid bottomonium state with $L=1$, $S=0$, $n=1$, and $J^{PC}=1^{--}$ then its radially excited state corresponding to $L=1$, $S=0$, $n=2$, and $J^{PC}=1^{--}$ has a predicted mass of $11014$ MeV, compatible with the experimentally observed mass of $\Upsilon(11020)$ i.e., $11000\pm4$ MeV \cite{PDG}. Nonetheless, in our previous work \cite{K2}, we have assigned $\Upsilon(11050)$ as $6\, {}^3S_1$ conventional bottomonium state and currently there are no experimental results disfavoring this assignment.

\begin{table}
	\caption{\label{C1} Mass spectra of $\Pi_u$ hybrid charmonium states in MeV.}
	\begin{ruledtabular}
		\begin{tabular}{@{}cccccccc}
			$L$ & $S$ & $J^{PC}$ & $n$ & Our work &\cite{3}&\cite{16}&\cite{37} \\
			\hline
			$1$&$0$&$1^{--}$&$1$&$4208$&$4212$&$4184(6)$&$4011$\\
			&&&$2$&$4465$&$4562$&$4572(10)$&$4355$\\
			&$1$&$0^{-+}$, $\mathbf{1^{-+}}$, $2^{-+}$&$1$&$4209$&$4212$&$4184(6)$&$4011$\\
			&&&$2$&$4466$&$4562$&$4572(10)$&$4355$\\
			$2$&$0$&$2^{++}$&$1$&$4356$&$4394$&$4374(8)$&$4231$\\
			&&&$2$&$4597$&&&$4563$\\
			&$1$&$1^{+-} $, $\mathbf{2^{+-}}$, $3^{+-}$&$1$&$4356$&$4394$&$4374(8)$&$4231$\\
			&&&$2$&$4597$&&&$4563$\\
			
		\end{tabular}
	\end{ruledtabular}
\end{table}

\begin{table}
	\caption{\label{C3} Mass spectra of $\Sigma_u^-$ hybrid charmonium states in MeV.}
	\begin{ruledtabular}
		\begin{tabular}{@{}cccccccc}
			$L$ & $S$ & $J^{PC}$ & $n$ & Our work&\cite{3}& \cite{16}&\cite{37} \\
			\hline
			$0$&$0$&$0^{++}$&$1$&$4544$&$4407$&$4487(5)$&$4486$\\
			&&&$2$&$4864$&$4822$&$4933(9)$&$4920$\\
			&$1$&$1^{+-}$&$1$&$4546$&$4407$&$4487(5)$&$4486$\\
			&&&$2$&$4867$&$4822$&$4933(9)$&$4920$\\
			$1$&$0$&$1^{--}$&$1$&$4673$&$4544$&$4623(6)$&$4011$\\
			&&&$2$&$4967$&&$5058(10)$&$4355$\\
			&$1$&$0^{-+}$, $\mathbf{1^{-+}}$, $2^{-+}$&$1$&$4673$&$4544$&$4623(6)$&$4011$\\
			&&&$2$&$4968$&&$5058(10)$&$4355$\\
			$2$&$0$&$2^{++}$&$1$&$4826$&&$4814(7)$&$4231$\\
			&&&$2$&$5089$&&&$4563$\\
			&$1$&$1^{+-}$, $\mathbf{2^{+-}}$, $3^{+-}$&$1$&$4826$&&$4814(7)$&$4231$\\
			&&&$2$&$5089$&&&$4563$\\
		\end{tabular}
	\end{ruledtabular}
\end{table}

\begin{table}
	\caption{Mass spectra of $\Pi_g$ hybrid charmonium states in MeV. \label{C2}}
	\begin{ruledtabular}
		\begin{tabular}{@{}ccccc}
			$L$ & $S$ & $J^{PC}$ & $n$ & Our work\\
			\hline
			$1$&$0$&$\mathbf{1^{-+}}$&$1$&$4593$\\
			&&&$2$&$4731$\\
			&$1$&$\mathbf{0^{--}}$, $1^{--}$, $2^{--}$&$1$&$4593$\\
			&&&$2$&$4732$\\
			$2$&$0$&$\mathbf{2^{+-}}$&$1$&$4664$\\
			&&&$2$&$4803$\\
			&$1$&$1^{++}$, $2^{++}$, $3^{++}$&$1$&$4664$\\
			&&&$2$&$4803$\\
		\end{tabular}
	\end{ruledtabular}
\end{table}

\begin{table}
	\caption{\label{C6} Mass spectra of $\Delta_g$ hybrid charmonium states in MeV.}
	\begin{ruledtabular}
		\begin{tabular}{@{}ccccc}
			$L$ & $S$ & $J^{PC}$ & $n$ &Our work \\
			\hline
			$2$&$0$&$2^{-+}$&$1$&$4694$\\
			&&&$2$&$4906$\\
			&$1$&$1^{--}$, $2^{--}$, $3^{--}$&$1$&$4694$\\
			&&&$2$&$4906$\\
		\end{tabular}
	\end{ruledtabular}
\end{table}

\begin{table}
	\caption{ Mass spectra of $\Sigma_g^-$ hybrid charmonium states in MeV. \label{C5}}
	\begin{ruledtabular}
		\begin{tabular}{@{}ccccc}
			$L$ & $S$ & $J^{PC}$ & $n$ & Our work\\
			\hline
			$0$&$0$&$\mathbf{0^{+-}}$&$1$&$5010$\\
			&&&$2$&$5262$\\
			&$1$&$1^{++}$&$1$&$5010$\\
			&&&$2$&$5262$\\
			$1$&$0$&$\mathbf{1^{-+}}$&$1$&$5065$\\
			&&&$2$&$5309$\\
			&$1$&$\mathbf{0^{--}}$, $1^{--}$, $2^{--}$&$1$&$5065$\\
			&&&$2$&$5309$\\
			$2$&$0$&$\mathbf{2^{+-}}$&$1$&$5156$\\
			&&&$2$&$5384$\\
			&$1$&$1^{++}$, $2^{++}$, $3^{++}$&$1$&$5156$\\
			&&&$2$&$5384$\\
		\end{tabular}
	\end{ruledtabular}
\end{table}

\begin{table}
	\caption{ Mass spectra of $\Delta_u$ hybrid charmonium states in MeV. \label{C7}}
	\begin{ruledtabular}
		\begin{tabular}{@{}ccccc}
			$L$ & $S$ & $J^{PC}$ & $n$ & Our work \\
			\hline
			$2$&$0$&$2^{--}$&$1$&$5023$\\
			&&&$2$&$5200$\\
			&$1$&$\mathbf{1^{-+}}$, $2^{-+}$, $\mathbf{3^{-+}}$&$1$&$5023$\\
			&&&$2$&$5200$\\
		\end{tabular}
	\end{ruledtabular}
\end{table}

\begin{table}
	\caption{ Mass spectra of $\Sigma_u^+$ hybrid charmonium states in MeV. \label{C4}}
	\begin{ruledtabular}
		\begin{tabular}{@{}ccccc}
			$L$ & $S$ & $J^{PC}$ & $n$ & Our work\\
			\hline
			$0$&$0$&$\mathbf{0^{--}}$&$1$&$5044$\\
			&&&$2$&$5258$\\
			&$1$&$\mathbf{1^{-+}}$&$1$&$5044$\\
			&&&$2$&$5258$\\
			$1$&$0$&$1^{++}$&$1$&$5086$\\
			&&&$2$&$5296$\\
			&$1$&$\mathbf{0^{+-}}$, $1^{+-}$, $\mathbf{2^{+-}}$&$1$&$5086$\\
			&&&$2$&$5296$\\
			$2$&$0$&$2^{--}$&$1$&$5157$\\
			&&&$2$&$5358$\\
			&$1$&$\mathbf{1^{-+}}$, $2^{-+}$, $\mathbf{3^{-+}}$&$1$&$5157$\\
			&&&$2$&$5358$\\
		\end{tabular}
	\end{ruledtabular}
\end{table}

\begin{table}
	\caption{\label{B1} Mass spectra of $\Pi_u$ hybrid bottomonium states in MeV.}
	\begin{ruledtabular}
		\begin{tabular}{@{}cccccccc}
			$L$ & $S$ & $J^{PC}$ & $n$ & Our work & \cite{3}& \cite{16}&\cite{37}\\
			\hline
			$1$&$0$&$1^{--}$&$1$&$10815$&$10559$&$10679(4)$&$10690$\\
			&&&$2$&$11014$&$10766$&$10899(6)$&$10885$\\
			&$1$&$0^{-+}$, $\mathbf{1^{-+}}$, $2^{-+}$&$1$&$10815$&$10559$&$10679(4)$&$10690$\\
			&&&$2$&$11014$&$10766$&$10899(6)$&$10885$\\
			$2$&$0$&$2^{++}$&$1$&$10917$&$10659$&$10783(5)$&$10819$\\
			&&&$2$&$11112$&&&$11005$\\
			&$1$&$1^{+-}$, $\mathbf{2^{+-}}$, $3^{+-}$&$1$&$10917$&$10659$&$10783(5)$&$10819$\\
			&&&$2$&$11112$&&&$11005$\\
			
		\end{tabular}
	\end{ruledtabular}
\end{table}

\begin{table}
	\caption{\label{B3} Mass spectra of $\Sigma_u^-$ hybrid bottomonium states in MeV.}
	\begin{ruledtabular}
		\begin{tabular}{@{}cccccccc}
			$L$ & $S$ & $J^{PC}$ & $n$ & Our work&\cite{3}& \cite{16}&\cite{37}\\
			\hline
			$0$&$0$&$0^{++}$&$1$&$11052$&$10892$&$10912(3)$&$11011$\\
			&&&$2$&$11320$&$11171$&$11192(5)$&$11299$\\
			&$1$&$1^{+-}$&$1$&$11052$&$10892$&$10912(3)$&$11011$\\
			&&&$2$&$11320$&$11171$&$11192(5)$&$11299$\\
			$1$&$0$&$1^{--}$&$1$&$11142$&$10979$&$10998\pm4$&$10690$\\
			&&&$2$&$11398$&&$11268\pm6$&$10885$\\
			&$1$&$0^{-+}$, $\mathbf{1^{-+}}$, $2^{-+}$&$1$&$11142$&$10979$&$10998\pm4$&$10690$\\
			&&&$2$&$11398$&&$11268\pm6$&$10885$\\
			$2$&$0$&$2^{++}$&$1$&$11262$&&$11117\pm4$&$10819$\\
			&&&$2$&$11502$&&&$11005$\\
			&$1$&$1^{+-}$, $\mathbf{2^{+-}}$, $3^{+-}$&$1$&$11262$&&$11117\pm4$&$10819$\\
			&&&$2$&$11502$&&&$11005$\\
		\end{tabular}
	\end{ruledtabular}
\end{table}

\begin{table}
	\caption{Mass spectra of $\Pi_g$ hybrid bottomonium states in MeV.\label{B2}}
	\begin{ruledtabular}
		\begin{tabular}{@{}ccccc}
			$L$ & $S$ & $J^{PC}$ & $n$ & Our work\\
			\hline
			$1$&$0$&$\mathbf{1^{-+}}$&$1$&$11308$\\
			&&&$2$&$11435$\\
			&$0$&$\mathbf{0^{--}}$, $1^{--}$, $2^{--}$&$1$&$11308$\\
			&&&$2$&$11435$\\
			$2$&$0$&$\mathbf{2^{+-}}$&$1$&$11385$\\
			&&&$2$&$11506$\\
			&$1$&$1^{++}$, $2^{++}$, $3^{++}$&$1$&$11385$\\
			&&&$2$&$11506$\\
		\end{tabular}
	\end{ruledtabular}
\end{table}

\begin{table}
	\caption{\label{B6} Mass spectra of $\Delta_g$ hybrid bottomonium states in MeV.}
	\begin{ruledtabular}
		\begin{tabular}{@{}ccccc}
			$L$ & $S$ & $J^{PC}$ & $n$ &Our work \\
			\hline
			$2$&$0$&$2^{-+}$&$1$&$11340$\\
			&&&$2$&$11503$\\
			&$1$&$1^{--}$, $2^{--}$, $3^{--}$&$1$&$11340$\\
			&&&$2$&$11503$\\
		\end{tabular}
	\end{ruledtabular}
\end{table}

\begin{table}
	\caption{ Mass spectra of $\Sigma_g^-$ hybrid bottomonium states in MeV. \label{B5}}
	\begin{ruledtabular}
		\begin{tabular}{@{}ccccc}
			$L$ & $S$ & $J^{PC}$ & $n$ &Our work\\
			\hline
			$0$&$0$&$\mathbf{0^{+-}}$&$1$&$11540$\\
			&&&$2$&$11749$\\
			&$1$&$1^{++}$&$1$&$11540$\\
			&&&$2$&$11749$\\
			$1$&$0$&$\mathbf{1^{-+}}$&$1$&$11582$\\
			&&&$2$&$11788$\\
			&$1$&$\mathbf{0^{--}}$, $1^{--}$, $2^{--}$&$1$&$11582$\\
			&&&$2$&$11788$\\
			$2$&$0$&$\mathbf{2^{+-}}$&$1$&$11651$\\
			&&&$2$&$11851$\\
			&$1$&$1^{++}$, $2^{++}$, $3^{++}$&$1$&$11651$\\
			&&&$2$&$11851$\\
		\end{tabular}
	\end{ruledtabular}
\end{table}

\begin{table}
	\caption{ Mass spectra of $\Sigma_u^+$ hybrid bottomonium states in MeV. \label{B4}}
	\begin{ruledtabular}
		\begin{tabular}{@{}ccccc}
			$L$ & $S$ & $J^{PC}$ & $n$ & Our work\\
			\hline
			$0$&$0$&$\mathbf{0^{--}}$&$1$&$11668$\\
			&&&$2$&$11835$\\
			&$1$&$\mathbf{1^{-+}}$&$1$&$11668$\\
			&&&$2$&$11835$\\
			$1$&$0$&$1^{++}$&$1$&$11698$\\
			&&&$2$&$11865$\\
			&$1$&$\mathbf{0^{+-}}$, $1^{+-}$, $\mathbf{2^{+-}}$&$1$&$11698$\\
			&&&$2$&$11865$\\
			$2$&$0$&$2^{--}$&$1$&$11750$\\
			&&&$2$&$11914$\\
			&$1$&$\mathbf{1^{-+}}$, $2^{-+}$, $\mathbf{3^{-+}}$&$1$&$11750$\\
			&&&$2$&$11914$\\
		\end{tabular}
	\end{ruledtabular}
\end{table}

\begin{table}
	\caption{ Mass spectraa of $\Delta_u$ hybrid bottomonium states in MeV. \label{B7}}
	\begin{ruledtabular}
		\begin{tabular}{@{}ccccc}
			$L$ & $S$ & $J^{PC}$ & $n$ & Our work \\
			\hline
			$2$&$0$&$2^{--}$&$1$&$11721$\\
			&&&$2$&$11858$\\
			&$1$&$\mathbf{1^{-+}}$, $2^{-+}$, $\mathbf{3^{-+}}$&$1$&$11721$\\
			&&&$2$&$11858$\\
		\end{tabular}
	\end{ruledtabular}
\end{table}

\section{Summary: \label{S4}}
In this work, we have computed the mass spectra of $\Pi_u$, $\Sigma_u^-$, $\Pi_g$, $\Delta_g$, $\Sigma_g^-$, $\Delta_u$, and $\Sigma_u^+$ hybrid charmonium and hybrid bottomonium states. The hybrid potential considered in this work is the sum of quark-antiquark potential and the gluonic excited potential. The quark-antiquark potential consists of short range one-gluon Coulomb potential, long range screened confinement potential, and $O(1/m)$ potential term. The gluonic excited potential term is obtained by fitting the difference between the hybrid potential and quark-antiquark potential computed from the lattice quantumchromodynamics. It is interesting to note that these differences for $\Pi_u$, $\Sigma_u^-$, $\Pi_g$, $\Delta_g$, $\Sigma_g^-$, $\Delta_u$, and $\Sigma_u^+$ hybrid states are observed to exhibit a common form $X+Ye^{-Zr}$. We predict the ground state mass of hybrid charmonium state as $4208$ MeV and hybrid bottomonium state as $10815$ MeV. From our model, the experimental mass of $\Psi(4230)$ is compatible with $\Pi_u$ hybrid charmonium state with $L=1$, $S=0$, $n=1$, and $J^{PC}=1^{--}$, $\Psi(4500)$ is compatible with a $\Pi_u$ hybrid charmonium state with $L=1$, $S=0$, $n=2$, and $J^{PC}=1^{--}$, $X(4500)$ is compatible with a $\Sigma_u^-$ hybrid charmonium state with $L=0$, $S=0$, $n=1$, and $J^{PC}=0^{++}$, $\Psi(4660)$ is compatible with a $\Sigma_u^-$ hybrid charmonium state with $L=1$, $S=0$, $n=1$, and $J^{PC}=1^{--}$, $\Upsilon(10753)$ is compatible with a $\Pi_u$ hybrid bottomonium state with $L=1$, $S=0$, $n=1$, and $J^{PC}=1^{--}$, and $\Upsilon(11050)$ is compatible with a  $\Pi_u$ hybrid bottomonium state with $L=1$, $S=0$, $n=2$, and $J^{PC}=1^{--}$. Our results suggest that the considered experimental states could serve as a potential candidates for hybrid mesons. Nevertheless, the conventional quarkonium interpretations cannot be ruled out for states with non-exotic quantum numbers using studies based solely on their spectra. Further theoretical and experimental studies on production mechanisms and decay properties are required to investigate the internal structure and to obtain a definitive conclusions.

\begin{acknowledgements}
	One of the authors (RK) is grateful to the Manipal Academy of Higher Education, Manipal for financial support under ‘Dr. TMA Pai Scholarship Programme’.
\end{acknowledgements}

\bibliographystyle{apsrev4-2}
\bibliography{ccc}

\end{document}